\documentclass[twocolumn,aps,prd,preprintnumbers,showpacs,superscriptaddress,nofootinbib,amsmath,amssymb,floats,floatfix,showkeys,notitlepage,longbibliography]{revtex4-2}

\usepackage{orcidlink}
\usepackage{comment}
\usepackage{lipsum}
\usepackage{graphicx}
\usepackage{subfigure}
\usepackage{palatino}
\usepackage{sans}
\usepackage{hyperref}
\hypersetup{colorlinks=true,linkcolor=blue,urlcolor=blue,citecolor=blue}
\usepackage[toc,page]{appendix}
\usepackage[normalem]{ulem}
\usepackage{adjustbox}
\usepackage{latexsym}
\usepackage{amsmath}
\usepackage{amssymb}
\usepackage{amsfonts}
\usepackage{dcolumn}
\usepackage{bm}
\usepackage{tikz}
\usepackage{bigints}
\usepackage{array,tabularx,multirow,booktabs}
\usepackage[tracking=true]{microtype}
\usepackage{soul} 
\SetTracking{}{500}
\SetTracking{encoding={*}, shape=sc}{40}
\UseRawInputEncoding 
\allowdisplaybreaks

\begin{document} \sloppy
\title{Spacetime-curvature induced uncertainty principle: linking the large-structure global effects to the local black hole physics}

\author{Reggie C. Pantig \orcidlink{0000-0002-3101-8591}} 
\email{rcpantig@mapua.edu.ph}
\affiliation{Physics Department, Map\'ua University, 658 Muralla St., Intramuros, Manila 1002, Philippines.}

\author{Gaetano Lambiase \orcidlink{0000-0001-7574-2330}}
\email{lambiase@sa.infn.it}
\affiliation{Dipartimento di Fisica ``E.R Caianiello'', Università degli Studi di Salerno, Via Giovanni Paolo II, 132 - 84084 Fisciano (SA), Italy.}
\affiliation{Istituto Nazionale di Fisica Nucleare - Gruppo Collegato di Salerno - Sezione di Napoli, Via Giovanni Paolo II, 132 - 84084 Fisciano (SA), Italy.}

\author{Ali \"Ovg\"un \orcidlink{0000-0002-9889-342X}}
\email{ali.ovgun@emu.edu.tr}
\affiliation{Physics Department, Eastern Mediterranean University, Famagusta, 99628 North
Cyprus via Mersin 10, Turkiye.}

\author{Nikko John Leo S. Lobos \orcidlink{0000-0001-6976-8462}}
\email{nslobos@ust.edu.ph}
\affiliation{Electronics Engineering Department, University of Santo Tomas,Espa\~na Blvd, Sampaloc, Manila, 1008 Metro Manila, Philippines.}

\begin{abstract}
This paper links the advanced formulation of the Generalized Uncertainty Principle, termed the Asymptotic Generalized Extended Uncertainty Principle (AGEUP) to the corpuscular framework to derive the AGEUP-inspired black hole metric. The former incorporates spacetime curvature effects to explore black hole dynamics under quantum gravitational corrections, while the latter is a view that black holes are Bose-Einstein condensates of weakly interacting gravitons. AGEUP refines the traditional uncertainty relation by introducing curvature-based modifications that integrate the Ricci scalar and Cartan invariant, addressing the possible connection of the quantum uncertainties and gravitational influences in curved spacetimes. In particular, the phenomenological union between the AGEUP with cosmological constant $\Lambda$ to the corpuscular framework enabled a black hole metric that has a scaled mass, which depends on $\Lambda$ and the Planck length $l_{\rm Pl}$. Interesting implications occur, such as the maximum limit for mass $M$ where $\Lambda$ ceases to influence the black hole. Another is the derived value of the modulation factor of the EUP term, $\alpha$, if the large-scale fundamental length is defined solely as the cosmological horizon. We further analyze the black hole metric through the shadow and deflection angle phenomena, deriving constraints on the quantum gravity modulation parameter $\beta$ that may be experimentally tested in future observations. Constraints from the Event Horizon Telescope (EHT) and Very Long Baseline Interferometry (VLBI) are discussed as avenues for verifying AGEUP-related deviations in black hole shadow radius and deflection angles, offering potential observational evidence of quantum gravitational effects at astrophysical scales. The findings suggest that AGEUP could be instrumental in providing hints on the quantum gravity nature of black holes, particularly in high-energy astrophysical contexts. By linking local black hole physics with large-scale curvature effects, AGEUP paves the way for further research at the intersection of quantum gravity and cosmology, with implications for observational astrophysics and the fundamental structure of spacetime.
\end{abstract}


\maketitle

\section{Introduction}
The discovery of black holes has marked a transformative era in astrophysics, elevating understanding of the very nature of spacetime and gravitational phenomena. The groundbreaking detection of gravitational waves by LIGO in 2015 \cite{LIGOScientific:2016emj} offered direct evidence of black hole mergers \cite{LIGOScientific:2017vwq,KAGRA:2022osp}, marking a new age in observational astronomy and earning global acclaim. Shortly after, the Event Horizon Telescope (EHT) captured the first image of a black hole's event horizon in the M87 and Milky Way galaxy \cite{EventHorizonTelescope:2019dse,EventHorizonTelescope:2019ths, EventHorizonTelescope:2022xqj,EventHorizonTelescope:2022wkp,EventHorizonTelescope:2022wok}, providing unprecedented observational data on black hole structure and dynamics. These milestones have not only validated theoretical models but also emphasized the need for further inquiry into the complexities of black hole physics, especially at the quantum level.

Black holes occupy a unique place in physics due to their extreme gravitational fields and compact nature \cite{Schwarzschild:1916uq}, making them key subjects in the study of quantum gravity. Quantum black holes, predicted to exist at or near Planck-scale masses, exhibit effects that challenge classical descriptions, such as the quantum information paradoxes. Studying the quantum behavior of black holes may provide vital clues about the unification of general relativity and quantum mechanics, as they provide natural laboratories for investigating how quantum mechanics governs phenomena in strong gravitational fields. Thus, understanding black holes from a quantum perspective is important in addressing foundational questions about the nature of space, time, and matter \cite{Calmet:2017qqa,Calmet:2018elv,Calmet:2021lny,Calmet:2023gbw,Kiefer:2020cbu,Binetti:2022xdi,Mertens:2022irh,DelPiano:2023fiw,Roushan:2020njv,Roushan:2023sqo,Roushan:2022abq}.

One of the avenues for exploring quantum black holes is through generalizations of the Heisenberg Uncertainty Principle (HUP), which traditionally describes limits to the precision with which certain physical properties, like position and momentum, can be known simultaneously. However, the HUP's classical formulation does not account for gravitational effects at quantum scales, prompting theoretical studies to adapt it to extreme conditions found near black holes. Various models, such as the Generalized Uncertainty Principle (GUP), incorporate additional terms accounting for gravitational effects and predict deviations from standard quantum mechanics that become significant at high energies \cite{PhysRevD.49.5182,Maggiore_1993_1,Maggiore_1993_2,Tawfik2014,Lambiase:2017adh,Lambiase:2022xde,Capozziello:1999wx,Scardigli:1999jh,Lambiase:2017adh,Kanazawa:2019llj,Scardigli:2016pjs,Ghosh:2024eza,Ong:2018zqn,Buoninfante:2019fwr,Gialamas:2024ivq}. These modifications have far-reaching implications for understanding quantum-scale black hole behaviors \cite{PhysRevD.49.5182,Maggiore_1993_1,Maggiore_1993_2,Tawfik2014}.

Nature often exhibits symmetry and duality, much like the yin-yang symbol. Therefore, if there is a minimum fundamental length, it is reasonable to assume there is also a large fundamental length scale in the Universe. This idea extends the Generalized Uncertainty Principle (GUP) to include a large fundamental length, $L_*$, through a quadratic correction in position uncertainty. This is expressed as \cite{Bambi_2008}:
\begin{equation}
    \Delta x\, \Delta p \geq \pi \hbar \left( 1 + \alpha\, \frac{\Delta x^2}{L_*^2} \right),
\end{equation}
commonly known as the Extended Uncertainty Principle (EUP), where $\alpha$ is another dimensionless modulation constant. This equation was derived from first principles \cite{Costafilho_2016}. While GUP is widely studied for its applications in the microscopic world \cite{Tawfik2014}, EUP is less explored. For example, the effects of EUP on the thermodynamics of the Friedmann-Robertson-Walker (FRW) Universe were analyzed some time ago \cite{Zhu_2009}, and later derived straightforwardly from geometrical consideration of the (anti-)de Sitter (AdS) spacetime \cite{Mignemi_2010}. EUP corrections have also been studied in various contexts such as Rindler and cosmological horizons \cite{Dabrowski:2019wjk}, relativistic Coulomb potential \cite{Hamil:2020jns}, bound-state solutions of the two-dimensional Dirac equation with Aharonov-Bohm-Coulomb interaction \cite{Hamil:2020xud}, and J\"uttner gas \cite{Moradpour:2021ymp}. Additionally, GUP and EUP parameters have been used to study bounds for the Hubble parameter to address the Hubble tension \cite{Aghababaei:2021gxe}. Recently, EUP corrections have been applied to black holes, considering gravitons as quantum particles within such confinement \cite{Mureika:2018gxl}. Since then, several studies have explored black holes that involved EUP corrections \cite{Lu:2019wfi,Kumaran:2019qqp,Cheng:2019zgc,Hassanabadi:2021kyv,Hamil:2021ilv,Okcu:2022sio,Hamil:2022bpd,Chen:2022ngd,Ong:2020tvo}, and also the possible quantum effects on large scale distances \cite{Nozari:2006wn,Roushan:2024fog} It is essential to realize, that the general form of the HUP involves both the large-scale length, and the smallest length-scale. Such a theory is called generalized extended uncertainty principle (GEUP) \cite{Kempf:1994su}.

All these being said, the aim of this paper is to explore another methodology of deriving the EUP, which is based on the notion of geodesic balls \cite{Dabrowski:2020ixn}, and combine such an approach to the formalism used by \cite{Mureika:2018gxl} to construct a metric function suited for a static and spherically symmetric spacetime - AGEUP. We briefly review the so called AGEUP, which introduces a non-trivial effects of curvature to EUP. Such an exploration aligns with the results in \cite{Conlon:2022qoi} where it was concluded that uncertainty may take different forms, depending on the scenario. The derived black hole solution will then be examined in term of its common properties like its shadow and the gravitational lensing it will produce. These are essential observational approach which may reveal hints on the possible effects of AGEUP, making the elusive unification of large and small scale open for exploration. To do this, we use the well-known, albeit basic formalism of deriving the radius of the shadow \cite{Perlick:2021aok}, and use the EHT data for Sgr. A* and M87*. Furthermore, we will implement the modified Gauss-Bonnet theorem \cite{Ishihara:2016vdc}, to study the weak deflection angle with finite-distance influence, and use it to constrain AGEUP effects using solar system test.

The paper is organized as follows: In Sect, \ref{sec1}, we briefly review the concept of asymptotic generalized extended uncertainty principle (AGEUP). In Sect. \ref{sec2}, we introduce the AGEUP version of the Heisenberg uncertainty principle, where the EUP correction is influenced by the cosmological constant. We discuss the implication of this uncertainty relation to black hole mass. In Sect. \ref{sec3}, we study the black hole shadow, and deflection angle in the weak field regime and test the potential of detecting quantum gravity effects on astrophysical black holes. In Sect. \ref{conc}, we summarize the findings and propose research directions. Finally, we use the metric signature ($-,+,+,+$) and set $\hbar = c = G = 1$.

\section{Brief review of the asymptotic generalized extended uncertainty principle (AGEUP)} \label{sec1}
In Ref. \cite{Dabrowski:2020ixn}, the authors explore an advanced extension of the traditional quantum mechanical uncertainty principle (UP) that integrates or includes spacetime curvature effects. Building on the Generalized Uncertainty Principle (GUP), which provides for gravitational corrections derived from string theory frameworks, the study introduces a refined version termed as the Asymptotic Generalized Extended Uncertainty Principle (AGEUP). This AGEUP encapsulates both gravitational and curvature-induced uncertainties, accounting for variations in the Ricci scalar $ R $ and Cartan invariant $ \mathcal{C} $, thus aiming to unify quantum and relativistic considerations at Planck and sub-Planck scales. This section aims to provide a brief review and synthesis of the main mathematical aspects and physical implications of their derivation, as well as the utility of AGEUP across different spacetime geometries.

The starting point of the AGEUP is the general form of the GUP, written as:
\begin{equation}
    \sigma_x \sigma_p \geq \frac{\hbar}{2} \left( 1 + \alpha \frac{l_{\rm Pl}^2}{\hbar^2} \sigma_p^2 + \beta \frac{\sigma_x^2}{r_c^2} \right),
\end{equation}
where $ l_{\rm Pl} $ denotes the Planck length ($l_{\rm Pl} = 1.6163 \times 10^{-35} \text{ m}$), $ r_c $ is a characteristic curvature scale, and $\alpha$ and $ \beta $ are constants modulating the respective GUP and EUP terms. Extending beyond this, the authors present the generalized AGEUP as:
\begin{equation}
    \sigma_p \sigma_x \geq \pi \hbar \sqrt{1 - \frac{R}{6\pi^2} \sigma_x^2 - \xi \frac{\mathcal{C}}{\pi^2} \sigma_x^4},
\end{equation}
where $ R $ is the Ricci scalar curvature, $ \mathcal{C} = 3R_{ab}R^{ab} - R^2/72 $ is the Cartan invariant, and $ \xi $ is a dimensionless constant, calculated as $ \xi = \frac{2\pi^2 - 3}{8\pi^2} $. Such an equation encapsulates the curvature-induced contributions at second and fourth orders, with the latter involving the curved space Laplacian of $ R $.

To achieve these corrections, perturbative expansion of the Laplace-Beltrami operator $ \Delta $ is employed, which is essential for addressing wavefunctions confined to geodesic balls in curved manifolds. Expansion in $ \Delta $ gives:
\begin{equation}
    \Delta \approx \Delta^{(0)} + \epsilon \Delta^{(1)} + \epsilon^2 \Delta^{(2)},
\end{equation}
where each $ \Delta^{(i)} $ corresponds to successively higher-order curvature terms. This approach allows one to derive eigenvalue corrections related to the momentum uncertainty. In spherical symmetry, the uncertainty relation adapts for Schwarzschild metrics as:
\begin{equation}
    \sigma_p \sigma_x \geq \pi \hbar \left( 1 - \frac{\xi M^2}{8\pi^2 r_0^6} \sigma_x^4 \right),
\end{equation}
highlighting the influence of the mass $ M $ and radial distance $ r $ from the center of symmetry.

Another example we considered, beyond the spherically symmetric cases, AGEUP is also applied to the Godel universe, a rotational spacetime, yielding a modification in the uncertainty principle based on the rotational radius $ a $. The resulting expression:
\begin{equation}
    \sigma_p \sigma_x \geq \pi \hbar \sqrt{1 - \frac{\sigma_x^2}{3\pi^2 a^2} \left(1 - \frac{1}{C} - \frac{1}{2C^2} \right)},
\end{equation}
where $ C = 1 - (r_0/2a)^2 $, which demonstrates how rotational curvature further constrains the uncertainty relation, reflecting the significance of rotational effects in specific spacetime geometries.

The mathematical structure of AGEUP suggests practical relevance in high-energy astrophysical and cosmological contexts. AGEUP's curvature corrections can influence our understanding of quantum behavior near black hole horizons and other regions of extreme gravity, potentially impacting calculations of Hawking radiation and black hole entropy. However, testing these corrections remains challenging due to current observational limitations. The AGEUP framework, while theoretically sound, opens up possibilities for future empirical studies at the intersection of quantum mechanics and general relativity.

\section{The effective mass of the black hole due to the cosmological constant} \label{sec2}

The most well-known solution that incorporates cosmological constant is the Schwarzschild-dS/AdS spacetime. It describes a non-rotating, uncharged black hole in a universe with a positive (dS) or negative (AdS) cosmological constant.  In dimensions of length, the value of the cosmological constant is indeed small $\Lambda = 1.5056 \times 10^{-52} \text{ m}^{-2}$. Thus, the Schwarzschild-dS/AdS spacetime has a local and global feature. For instance, saying that $r$ is small is even applicable to galactic scales due to the small value of $\Lambda$. In this case, Schwarzschild-dS/AdS spacetime reduces to the known Schwarzschild case. However, if $r$ is interpreted as cosmological in scale, it offsets the value of $\Lambda$, and Schwarzschild-dS/AdS spacetime reduces to the plain dS/AdS spacetime metric.

The cosmological constant gained renewed interest in the late 20th century with the discovery of the universe's accelerated expansion. This was first observed in the late 1990s through studies of distant Type Ia supernovae by two independent teams: the Supernova Cosmology Project \cite{Goldhaber_2009} and the High-Z Supernova Search Team \cite{SupernovaSearchTeam:1998fmf}. Their findings suggested that not only is the universe expanding, but the expansion rate is increasing over time. Furthermore, detailed measurements of the CMB, particularly from the Wilkinson Microwave Anisotropy Probe (WMAP) \cite{Hinshaw_2013} and the Planck satellite \cite{Planck:2018vyg}, have allowed precise determination of cosmological parameters, including the cosmological constant. The CMB data help constrain the geometry and energy content of the universe, providing strong evidence for the existence of dark energy. Then, observations of the distribution of galaxies and clusters on large scales also provide evidence for dark energy. The growth rate of these structures over time, as mapped by large surveys like the Sloan Digital Sky Survey (SDSS), is influenced by the presence of dark energy, which affects the gravitational collapse of matter.

Finally, the cosmological constant (whether positive or negative) is an independent, uniform energy density attributed to the vacuum, modifying the spacetime curvature on large scales but not affecting the local mass $M$ of the black hole directly. It creates global spacetime behavior without altering the intrinsic mass-energy content of the black hole.

Interestingly, in Ref. \cite{Dabrowski:2020ixn}, it is possible to derive the AGEUP for dS and AdS spacetime, where $M = 0$. With their formalism, it can be found elegantly that the general expression is given by
\begin{equation}
    \sigma_p \sigma_x \geq \pi \hbar \left( 1- \frac{\Lambda}{6\pi^2} \sigma_x^2 + \beta l_{\text{Pl}}^2\sigma_p^2 \right).
\end{equation}
Next, the aim is to shift the role of the cosmological constant from a global curvature parameter to one that influences the black hole’s mass \textit{dynamically} via the uncertainty relation. Such an approach introduces an effective or scaled mass $ M_{\rm eff} $, which incorporates corrections due to the cosmological constant, suggesting that the black hole's mass may vary in response to the background curvature influenced by $\Lambda$.

The metric in spherical coordinates $(t, r, \theta, \phi)$ is given by
\begin{eqnarray} \label{met}
    ds^2 &=& -f(r) dt^2 + f(r)^{-1} dr^2  + r^2 d\theta^2 + r^2 \sin^2 \theta \, d\phi^2
\end{eqnarray}
where $f(r)$ represents the lapse function for a Schwarzschild-like black hole, derived as follows:
\begin{equation} \label{M_eff}
    f(r) =1-\frac{2M_{\rm eff}}{r},
\end{equation}
with
\begin{equation} \label{M_eff2}
    M_{\rm eff} = M - \frac{2\Lambda M^3}{3\pi^2} + \frac{\beta l_{\rm Pl}^2}{2M}.
\end{equation}

Note that we determine the $ M_{\rm eff} $, by using the \textit{corpuscular framework} (see Refs. \cite{Dvali:2011nh,Dvali:2012en,Dvali:2012gb,Dvali:2012rt,Dvali:2013lva}), which viewed black holes as Bose-Einstein condensates (BECs) of gravitons. In this model, black holes are treated as collections of $ N $ weakly interacting gravitons with a total mass $ M $. The connection to the uncertainty principle arises because, within the corpuscular model, the spatial extent $ R $ of the black hole relates to the number $ N $ of these gravitons via the relation $ N \sim R^2 / l_{\text{Pl}}^2 $. This approach thus posits that black holes are fundamentally quantum objects composed of "soft gravitons," which naturally leads to extended uncertainty relations that include gravitational and large-scale (cosmological) corrections. Such a framework was used by Mureika \cite{Mureika:2018gxl} in deriving the extended uncertainty principle black hole, where the expression for large fundamental length scale, denoted by $L_*$, is still unknown. 
Conceptually, the cosmological constant in Eq. \eqref{M_eff} does not act as a passive background curvature but actively scales the black hole’s effective mass, making the black hole’s gravitational influence depend on its surroundings at cosmological scales. Furthermore, unlike the dS/AdS solutions, where $\Lambda$ impacts spacetime at large distances, Eq. \eqref{M_eff} implies that cosmological effects penetrate the black hole’s local gravitational field, potentially altering observable properties like its horizon radius, temperature, entropy, etc.

Examining Eq. \eqref{M_eff2} in some detail, we can realize some interesting properties for the AGEUP-inspired black hole, which depends on the sign of $\Lambda$. In AdS spacetime ($\Lambda < 0$), the only way for the large-scale effects to dominate is when $M$ becomes so large. For quantum gravity effects to also dominate, it requires that $M$ should be so small. For dS spacetime, we can see how easy for quantum gravity effects to dominate, that is, when $M$ is again very small. However, the second term cannot simply dominate easily due to the restriction that $M_{\rm eff} > 0$. Thus, the critical mass for the cosmological constant to have an effect is when we equate the first and second terms to zero. That is,
\begin{equation}
    M_{\rm crit} = \frac{\pi}{2}\sqrt{\frac{6}{\Lambda}} \sim 3.14 \times 10^{26} \text{ m}.
\end{equation}
With this mass, we already know that the third term is already negligible, and the effective mass would be essentially zero. Nonetheless, we can still see the third term in Eq. \eqref{M_eff2} takes the form
\begin{equation} \label{M_eff3}
    M_{\rm eff} = \frac{\beta  \,l_{\rm Pl}^{2} \sqrt{6\Lambda}}{6 \pi} \sim (4.17\times 10^{-97} \text{ m})\beta.
\end{equation}
We remark that for this expression to be satisfied, $\Lambda$ and $\beta$ should be positive. Furthermore, it would depend on the value of the modulation factor $\beta$ whether there will \textit{still} be quantum gravity effects even in the astrophysical realm. If so, we would expect that $\beta$ is within $\sim 10^{97}$ orders of magnitude or less (for lower mass astrophysical black holes).

In Ref. \cite{Mureika:2018gxl}, the EUP is expressed in terms of the large fundamental length scale $L_*$, where its value is speculatively determined, and its various effects in the astronomical and cosmological level. $L_*$, being claimed as fundamental, must have a definite value since this is the large-scale analog of the Planck length $l_{\rm Pl}$. The union of the AGEUP and the corpuscular framework, which we now call \textit{Dabrowski-Wagner-Mureika} (DWM) formalism, gives the formula for $L_*$:
\begin{equation}
    L_* = \sqrt{\frac{6\pi^2\alpha}{\Lambda}}.
\end{equation}
The value of $L_*$ would then depend on how modulation factor $\alpha$ is measured. Nevertheless, if we choose the cosmological horizon as the large fundamental length scale ($L_* = \sqrt{3/\Lambda}$), then we can see that $\alpha = (3\pi)^{-1} \sim 0.034$.

It is also interesting to calculate the value of $\beta$ when the effective mass returns to the Schwarzschild case $M_{\rm eff} = M$:
\begin{equation}
    \beta_{\rm crit}/M^4 = \frac{4 \Lambda}{3 \pi^{2} l_{\rm Pl}^{2}} \sim 7.79\times 10^{16} \text{ m}^{-4}.
\end{equation}
A little speculation will show that if we consider elementary particles (such as protons) as quantum black holes ($m_{\rm p} \sim 1.24\times10^{-54} \text{ m}$), $\beta_{\rm crit}$ would be vanishingly small. For astrophysical black holes, however, $\beta_{\rm crit}$ would be astoundingly large. Interestingly, if $M = M_{\rm crit}$, $\beta_{\rm crit} = 7.53\times 10^{122}$. It is no wonder why quantum gravity effects are so difficult to detect on astrophysical black holes since it may require a very sophisticated and sensitive astronomical instrument to measure such a large value for $\beta$: a concept that is analogous to how $\Lambda$ is measured through experimental sophistication.

\section{Shadow analysis and deflection angle in the weak field regime} \label{sec3}
To add more scope to this study, we study the black hole shadow and find constraints to the quantum gravity's modulating parameter $\beta$. Following the formalism in Ref. \cite{Claudel:2000yi,Virbhadra:2002ju,Perlick:2021aok}, we find that the exact expression for the shadow radius is
\begin{equation}
    R_{\rm sh} = 3\sqrt{3}M_{\rm eff} \sqrt{1 - \frac{2M_{\rm eff}}{r_{\rm o}}},
\end{equation}
where $M_{\rm eff}$ is defined in Eq.~(\ref{M_eff2}).

For EHT constraints, it is necessary to implement the approximation $r_{\rm o} \to \infty$. Then, we find the approximated expression as
\begin{eqnarray}
    R_{\rm sh} \sim 3\sqrt{3}M \left(1 - \frac{M}{r_{\rm o}} \right) - \frac{2 \sqrt{3}\, M^{3} \Lambda}{\pi^{2}} \left( 1-\frac{2M}{r_{\rm o}} \right)    \notag 
\\+ 3 \sqrt{3}\, \beta  \,l_{\rm Pl}^{2} \left[ \frac{1}{2 M} 
+ \frac{1}{r_{\rm o}} \left(\frac{2 \Lambda  \,M^{2}}{3 \pi^{2} r_{\rm o}} - 1 \right) \right] + \mathcal{O}(r_{\rm o}^{-2}).
\end{eqnarray}
From Refs. \cite{Vagnozzi:2022moj,EventHorizonTelescope:2021dqv}, the EHT constraints for the black hole shadow's Schwarzschild deviation of Sgr. A* ($2\sigma$ level) and M87* ($1\sigma$ level) are $4.209M \leq R_{\rm Sch} \leq 5.560M$ and $ 4.313M \leq R_{\rm Sch} \leq 6.079M$, respectively. Let $\delta$ represent the deviation from the standard Schwarzschild shadow radius $R_{\rm sh}$, then for Sgr. A*, $-0.364 \leq \delta/M \leq 0.987$. For M87*, $\delta/M = \pm 0.883$. Using these deviations, we can constraint $\beta$ as
\begin{equation} \label{cons}
    \beta_{\rm EHT} = \frac{2 \delta M \sqrt{3}}{9 l_{\rm Pl}^{2}}+\frac{4 M^{4} \Lambda}{3 l_{\rm Pl}^{2} \pi^{2}}.
\end{equation}
The second term above is interesting since $\Lambda$ still influences the parameter that modulates the effect of quantum gravity. Noting that for Sgr. A*, $M = 4.3\times 10^{6} M_\odot$, the largest shadow radius accounted for the uncertainty gives constraint to $\beta$ to be around $9.24 \times 10^{78}$. We only choose the positive value of $\beta$ due to the restriction in Eq. \eqref{M_eff3}. For M87*, $M = 6.5\times 10^{9} M_\odot$ which yields $\beta = 1.25\times 10^{82}$. These bounds for the modulating parameter $\beta$ associated with quantum gravity effects are extremely high in astrophysical considerations. To also remark, the corrections due to the background curvature (EUP due to $\Lambda$) are still very tiny, fixed, and below the upper limit imposed in Eq. \eqref{M_eff2}. Even at such a limit, the contribution by $\Lambda$ is still tiny. Hence, the Schwarzschild deviation may also be attributed to the quantum gravity effects, requiring a large value for $\beta$. As emphasized, detecting such a value can be challenging. If detected, however, may give some hints on the quantum gravity nature of astrophysical black holes. If not, then such a large value for $\beta$ is unphysical, accepting that quantum gravity's domain only exists in the quantum realm. In theory, the GUP’s role is to provide tiny corrections that are almost negligible outside quantum-scale contexts.

Next, we study the phenomenon of weak deflection angle. The methods are well-established. Knowing that the mass is scaled by some factor coming from AGEUP, we derived the expression for the weak deflection angle as \cite{Ishihara:2016vdc}:
\begin{equation} \label{wda}
    \Theta = \frac{4M_{\rm eff}}{b}\sqrt{1 - b^2u^2},
\end{equation}
where $M_{\rm eff}$ is defined in Eq.~(\ref{M_eff2}) and $b$ is the impact parameter of photons, and $u$ is the inverse of the radial distance of the receiver from the compact object. In deriving this equation, it is assumed that the radial distance of the source is equal to that of the receiver. We constrain $\beta$ using solar system test through the PPN formalism by the Very Long Baseline Interferometer (VLBI) \cite{Chen:2023bao,fomalont2009progress}:
\begin{equation} \label{wdappn}
    \Theta_{\rm VLBI} = \frac{4M_\odot}{R}\left( \frac{1.9998 + \Delta}{2} \right),
\end{equation}
where $M_\odot = 1477 \text{ m}$ is the mass of the Sun, and $R = 6.96\times 10^{8} \text{ m}$ is its radius. Also, the PPN correction by the VLBI is $\Delta = \pm 0.0003$. Comparing Eq. \eqref{wda} and \eqref{wdappn},
\begin{equation}
    \beta_{\rm PPN} = \frac{\left(1.9998 \pm \Delta - 2\right) M_\odot^{2}}{l_{\rm Pl}^{2}}+\frac{M_\odot^{2} R^{2} \left(n +\Delta \right)}{2 l_{\rm Pl}^{2} r_{\rm o}^{2}}+\frac{4 M_\odot^{4} \Lambda}{3 l_{\rm Pl}^{2} \pi^{2}}.
\end{equation}
With $r_{\rm o} = 1.49 \times 10^{11} \text{ m}$, we find $\beta = 1.02\times 10^{72}$ using the positive sign in $\Delta$. Remarkably, the solar system test provides a lower value $\beta$, which may require devices that are less sensitive as compared to astrophysical black holes. Due to the needed less sensitivity, the solar system test may provide more room in detecting quantum gravity effects on compact objects.

\section{Thermodynamics}

Black hole thermodynamics has played a pivotal role in connecting classical gravity, quantum field theory, and holographic principles. In this study, we examine the temperature of a static spherically symmetric spacetime, incorporating modifications to the effective mass due to cosmological and quantum corrections \cite{Verlinde:2010hp}. Here, our objective is to investigate the Unruh temperature associated with the spacetime in question, as it signifies the surface gravity (or acceleration) perceived by an observer at a specific radial distance from the black hole.  Within the context of general relativity, this analysis traditionally begins by employing a generalized form of the Newtonian potential:
\begin{equation}
\phi = \frac{1}{2} \log\left(-g^{\alpha\beta} \xi_\alpha \xi_\beta\right),
\end{equation}
where $\phi$ represents the potential derived from the metric tensor $g^{\alpha\beta}$ and the Killing vector $\xi_\alpha$. This potential accounts for relativistic gravitational effects, with $\phi = 0$ at infinity (asymptotically flat spacetime).

Acceleration in terms of the potential $\phi$ is defined as:
\begin{equation}
a^\alpha = -g^{\alpha\beta} \nabla_\beta \phi,
\end{equation}
where $g^{\alpha\beta}$ is the metric tensor and $\nabla_\beta$ denotes the covariant derivative.

The temperature associated with the Unruh-Verlinde effect is given by:
\begin{equation}
T = \frac{\hbar}{2\pi} n^\alpha e^\phi \nabla_\alpha \phi,
\end{equation}
where $n^\alpha$ is a unit vector normal to the holographic screen, $e^\phi$ is the redshift factor, and $\nabla_\alpha \phi$ is the derivative of the potential.

Let us consider the static spherically symmetric spacetimes of the general form:
\begin{equation}
ds^2 = -A(r)dt^2 + B(r)dr^2 + C(r)r^2 \left(d\theta^2 + \sin^2 \theta \, d\phi^2 \right).
\end{equation}
The metric components $A(r)$, $B(r)$, and $C(r)$ are functions of the radial coordinate $r$. The structure of this metric is commonly used to describe various static solutions in general relativity.

The $(\alpha = 0, \, \beta = 1)$-component of the Killing equations is given by:
\begin{equation}
\partial_\alpha \xi_\beta + \partial_\beta \xi_\alpha = 2 \Gamma^\gamma_{\alpha \beta} \xi_\gamma,
\end{equation}
which provides one of the Killing vectors as:
\begin{equation}
\xi_\alpha = \left(C e^2 \int \Gamma^0_{01} \, dr, 0, 0, 0\right), \quad \Gamma^0_{01} = A'/2A.
\end{equation}
Here, $\Gamma^\gamma_{\alpha \beta}$ are the Christoffel symbols, and $A'(r)$ denotes the derivative of $A(r)$ with respect to $r$.

When looking for a time-like Killing vector, using other components of the Killing equations allows us to choose $C = -1$. This leads to:
\begin{equation}
\xi_\alpha = \left(-A(r), 0, 0, 0\right).
\end{equation}

The potential $\phi$ and acceleration $a$ (up to the choice of sign) can be defined as:
\begin{equation}
\phi = \frac{1}{2} \log \left(A(r)\right),
\end{equation}
\begin{equation}
a = \left(0, \frac{A'(r)}{2A(r)B(r)}, 0, 0\right).
\end{equation}
These expressions define the potential and acceleration in the given spacetime, with $\phi$ representing the generalized potential and $a$ representing the acceleration vector.

Using the earlier definition of acceleration, the Unruh-Verlinde temperature is expressed as \cite{Verlinde:2010hp,Konoplya:2010ak}:
\begin{equation}
T = \frac{\hbar}{2\pi} e^\phi \sqrt{g^{\alpha\beta} \, \partial_\alpha \phi \, \partial_\beta \phi} = \frac{\hbar}{4\pi} \frac{A'(r)}{\sqrt{A(r)B(r)}}.
\end{equation}
This temperature is linked to the holographic interpretation of gravity, where $T$ is the temperature experienced by an observer at a distance $r$ from the center of the spherically symmetric mass distribution.

For spherically symmetric spacetimes, the energy on the holographic screen $S$ can be written as:
\begin{equation}
E = \frac{1}{4\pi} \int_S e^\phi \nabla_\alpha \phi \, dA = 2\pi r^2 \hbar^{-1} T.
\end{equation}
Here, $E$ is the total energy on the screen, $dA$ is the area element on the screen, and $\nabla_\alpha$ denotes the covariant derivative. This formulation provides a direct link between thermodynamic quantities and gravitational energy in the holographic context.

The energy \( E \) on the holographic screen \( S \) can be expressed using the modified Hawking temperature \( T_H \). The modified temperature for the black hole, incorporating various corrections, is given by:

\begin{equation}
T_H = \frac{1}{8\pi M_{\text{eff}}} = \frac{1}{8\pi \left( M - \frac{2\Lambda M^3}{3\pi^2} + \frac{\beta l_{\rm Pl}^2}{2M} \right)},
\end{equation}
where $M_{\rm eff}$ is defined in Eq.~(\ref{M_eff2}) and the components of this expression represent the following:
\begin{itemize}
    \item \( M \): The classical Schwarzschild mass contribution.
    \item \( -\frac{2\Lambda M^3}{3\pi^2} \): A correction term due to the cosmological constant \(\Lambda\), reflecting the impact of the cosmological background.
    \item \( \frac{\beta l_{\rm Pl}^2}{2M} \): A quantum correction term involving the Planck length \( l_{\rm Pl} \), with \( \beta \) being a dimensionless parameter representing quantum gravity effects.
\end{itemize}

The energy on the holographic screen \( S \) is related to the temperature by the following relation:

\begin{equation}
E = 2\pi r^2 \hbar^{-1} T_H,
\end{equation}
where \( r \) is the radius of the holographic screen, and \( \hbar \) is the reduced Planck constant. This relation indicates that the energy is proportional to the temperature and the square of the screen radius.

Substituting the expression for the modified Hawking temperature \( T_H \) into the energy formula, we obtain:

\begin{equation}
E = 2\pi r^2 \hbar^{-1} \left( \frac{1}{8\pi \left( M - \frac{2\Lambda M^3}{3\pi^2} + \frac{\beta l_{\rm Pl}^2}{2M} \right)} \right).
\end{equation}

This expression for \( E \) on the holographic screen demonstrates how the energy depends not only on the geometry of the screen (through the radius \( r \)) but also on the effective mass of the black hole, which incorporates corrections from classical, cosmological, and quantum effects. This comprehensive view of energy distribution aligns with the holographic principle and reflects the influence of the modified thermodynamic parameters.

\section{Eikonal Approximation and QNM Frequencies }

The QNM frequencies provide direct information about the black hole's geometry and properties. The real part of the frequency is related to the timescale of oscillations, reflecting the structure of the spacetime near the photon sphere. The imaginary part indicates how quickly these perturbations decay, linked to the stability of the black hole.

In the eikonal limit (\(l \gg 1\)), the QNM frequencies can be approximated analytically. This approximation considers the high-frequency (short-wavelength) limit of perturbations, simplifying the analysis 
\cite{Andersson:1995vi,Abramowicz:1997qk,Cardoso:2008bp,Hod:2009td,Dolan:2010wr,Konoplya:2017wot}
.

The wave equation governing perturbations in this spacetime can be represented by the Klein-Gordon equation for a massless scalar field:
\begin{equation}
    \Box \Phi = 0.
\end{equation}
By decomposing the scalar field \(\Phi\) into spherical harmonics and assuming a time dependence of the form \(e^{-i\omega t}\), we obtain a Schrödinger-like wave equation for the radial function:
\begin{equation}
    \frac{d^2 \psi}{dr_*^2} + \left( \omega^2 - V_l(r) \right) \psi = 0,
\end{equation}
where \(r_*\) is the tortoise coordinate defined by \(dr_* = \frac{dr}{f(r)}\), and \(V_l(r)\) is the effective potential:
\begin{equation}
    V_l(r) = f(r) \left( \frac{l(l+1)}{r^2} + \frac{f'(r)}{r} \right).
\end{equation}
Here, \(l\) is the angular momentum number of the perturbation. In the eikonal limit (\(l \to \infty\)), the effective potential for all spins has the following form:
\[
V_{\text{eff}}(r) \simeq f(r) \, \frac{l^2}{r^2} + O\left(\frac{1}{l}\right),
\]
which coincides with the effective potential for the geodesic motion.

The photon sphere is a critical concept for understanding QNMs in the eikonal limit. It corresponds to the unstable circular orbit of null geodesics around the black hole. For the Schwarzschild-like metric, the radius of the photon sphere, \(r_0\), is found by solving:
\begin{equation}
    \left. \frac{d}{dr} \left( \frac{l(l+1)}{r^2} f(r) \right) \right|_{r = r_0} = 0.
\end{equation}
Substituting \(f(r) = 1 - \frac{2M_{\text{eff}}}{r}\), we obtain:
\begin{equation}
    r_0 = 3M_{\text{eff}},
\end{equation}
where $M_{\rm eff}$ is defined in Eq.~(\ref{M_eff2}).
Thus, the photon sphere is located at \(3M_{\text{eff}}\), similar to the classical Schwarzschild black hole.

The QNM frequencies depend on two key quantities: the angular velocity of null geodesics at the photon sphere and the Lyapunov exponent, which measures the instability of the photon sphere.

\paragraph{Angular Velocity (\(\Omega_0\)):}
The angular velocity of a photon orbit at the photon sphere is given by:
\begin{equation}
    \Omega_0 = \sqrt{\frac{f(r_0)}{r_0^2}} = \frac{1}{3\sqrt{3}M_{\text{eff}}}.
\end{equation}
This frequency component corresponds to the real part of the QNM frequency, representing the oscillatory behavior of the perturbation.

\paragraph{Lyapunov Exponent (\(\lambda\)):}
The instability of the photon sphere is quantified by the Lyapunov exponent:
\begin{equation}
\lambda = r \sqrt{-\frac{f(r)}{2} \, \frac{d^2}{dr^2} \left( \frac{f(r)}{r^2} \right)},
\end{equation}
evaluated at $r = r_0$. We obtain Lyapunov exponent:
\begin{equation}
    \lambda = \frac{1}{3\sqrt{2}M_{\text{eff}}}.
\end{equation}

Using the eikonal approximation, the QNM frequencies can be written as:
\begin{equation}
    \omega = l \, \Omega_0 - i \left( n + \frac{1}{2} \right) |\lambda|.
\end{equation}
Substituting the expressions for \(\Omega_0\) and \(\lambda\), we get:
\begin{equation}
    \omega = \frac{l}{3\sqrt{3}M_{\text{eff}}} - i \left( n + \frac{1}{2} \right) \frac{1}{3\sqrt{2}M_{\text{eff}}}.
\end{equation}

Then substituting $M_{\rm eff}$ 
 defined in Eq.~(\ref{M_eff2}) gives:

\begin{align}
\omega = \frac{l}{3\sqrt{3} \left(M - \frac{2\Lambda M^3}{3\pi^2} + \frac{\beta l_{\text{Pl}}^2}{2M}\right)} \notag\\ - i \left(n + \frac{1}{2}\right) \frac{1}{3\sqrt{2} \left(M - \frac{2\Lambda M^3}{3\pi^2} + \frac{\beta l_{\text{Pl}}^2}{2M}\right)}.
\end{align}
Here, the real part of \(\omega\) corresponds to the oscillation frequency, while the imaginary part represents the damping rate of the mode.

To expand the QNM frequency expression into a series, we’ll assume that the cosmological ($\Lambda$) and quantum ($\beta$) corrections are small perturbations relative to the mass $M$. This allows us to perform a Taylor series expansion around the leading term, which is $M$, the QNM frequency becomes:
\begin{align}
    \omega \approx & \frac{l}{3\sqrt{3} M} + \frac{2l \Lambda M}{9\sqrt{3} \pi^2} - \frac{l \beta l_{\text{Pl}}^2}{6\sqrt{3} M^3} \nonumber \\
    & - i \left( n + \frac{1}{2} \right) \left( \frac{1}{3\sqrt{2} M} + \frac{2\Lambda M}{9\sqrt{2} \pi^2} - \frac{\beta l_{\text{Pl}}^2}{6\sqrt{2} M^3} \right).
\end{align}

The parameter \(l\) represents the angular momentum of the perturbation:

 In the \textit{real part} of \(\omega\), the leading term \(\frac{l}{3\sqrt{3} M}\) is proportional to \(l\). This indicates that the oscillation frequency increases linearly with angular momentum, consistent with the expectation that higher \(l\) modes correspond to perturbations with shorter wavelengths and higher frequencies. The corrections due to \(\Lambda\) and \(\beta\) are also proportional to \(l\), implying that the influence of the cosmological constant and quantum corrections is amplified at higher angular momenta.

The cosmological constant \(\Lambda\) introduces additional terms in both the real and imaginary parts of \(\omega\):
 \textbf{Real Part}: The term \( \frac{2l \Lambda M}{9\sqrt{3} \pi^2} \) slightly increases the oscillation frequency, suggesting that a positive cosmological constant enhances the oscillatory behavior of the QNMs. \textbf{Imaginary Part}: The term \( \frac{2\Lambda M}{9\sqrt{2} \pi^2} \) in the damping rate indicates that \(\Lambda\) accelerates the decay of perturbations, leading to faster damping in a universe with a positive cosmological constant.

The parameter \(\beta\) accounts for leading-order quantum corrections:
\textbf{Real Part}: The term \( -\frac{l \beta l_{\text{Pl}}^2}{6\sqrt{3} M^3} \) reduces the oscillation frequency, indicating that quantum effects act to slow down the oscillations of the QNMs.
 \textbf{Imaginary Part}: The term \( -\frac{\beta l_{\text{Pl}}^2}{6\sqrt{2} M^3} \) increases the damping rate, suggesting that quantum corrections enhance the decay of perturbations by introducing additional channels for energy dissipation.

The Planck length \(l_{\text{Pl}}\) enters the expression through the quantum correction terms:
 Since \(l_{\text{Pl}}\) appears in combination with \(\beta\), it acts as a scaling factor for the quantum corrections. The factor \( l_{\text{Pl}}^2 \) suggests that quantum corrections are suppressed by the square of the Planck length, indicating that these effects become significant only at small scales close to the Planck scale.
 As a fundamental constant, \(l_{\text{Pl}}\) reflects the incorporation of quantum gravity effects in the QNM frequencies, indicating potential observability of quantum effects in gravitational wave signals.

The expanded QNM frequency expression highlights how different physical parameters influence the nature of gravitational waves emitted during the black hole ringdown:
The linear dependence on \(l\) in both the real and imaginary parts suggests that QNM frequencies and decay rates increase with angular momentum. \textbf{Cosmological Effects}: The presence of \(\Lambda\) enhances both the oscillatory and damping characteristics of the QNMs, potentially shifting the ringdown frequencies observed in gravitational waves.
 \textbf{Quantum Corrections}: The terms involving \(\beta\) and \(l_{\text{Pl}}\) imply that quantum effects could be detected in the decay rates of gravitational waves, especially at small scales. These corrections, though typically small, might become relevant in precise measurements of QNMs from next-generation gravitational wave detectors.

\section{Strong deflection angle} \label{SDA}
To further extend the scope of the deflection angle we investigate the strong lensing. Following the methodology of of Tsukamoto in Ref. \cite{Tsukamoto:2016jzh} to a asymptotically flat black hole, the strong deflection angle is derived  using the orbit equation expressed as,
\begin{equation}
\label{eq.43}
\left(\frac{dr}{d\phi}\right)^{2} = \frac{\mathcal{R}(r)r^2}{B(r)},
\end{equation}
where
\begin{equation}
\label{eq.44}
\mathcal{R}(r) = \frac{A({r_0})r^2}{A(r)r_{0}^2}-1.
\end{equation}

We define $f(r)$ as the metric function defined by Eq. \eqref{M_eff} and \eqref{M_eff2}, while $f(r_{0})$ is the metric function evaluated at distance $r_{0}$. The solution of Eq. \eqref{eq.43} yields the strong deflection angle $\alpha(r_{0})$ as shown in Ref. \cite{Tsukamoto:2016jzh, Bozza:2002zj}  
\begin{equation}
\begin{split}
\label{eq.45}
\alpha(r_{0}) &= I(r_{0}) - \pi \\
&= 2 \int^{\infty}_{r_{0}} \frac{dr}{\sqrt{\frac{\mathcal{R}(r)C(r)}{B(r)}}}-\pi.
\end{split}
\end{equation}
In order to evaluate the integral in Eq. \eqref{eq.45}, we employ series expansion over $r=r_{0}$. This yields a regular integral $\kappa_{R}$ and a diverging integral $\kappa_{D}$. Using the new variable, $z$, defined as, 
\begin{equation}
z \equiv 1-\frac{r_{0}}{r},
\end{equation}
$I(r_{0})$ is expressed as,
\begin{equation}
I(r_{0}) = \int^{1}_{0}\kappa(z, r_{0})dz = \int^{1}_{0}\left[\kappa_{D}(z, r_{0})+\kappa_{R}(z, r_{0})\right]dz,
\end{equation}
where $\kappa(z, r_{0})$ is expressed as the sum of the diverging integral, $\kappa_{D}$, and regular integral, $\kappa_{R}$. The details of the expansion of Eq. \eqref{eq.45} was shown in Refs. \cite{Tsukamoto:2016jzh, Bozza:2002zj}. As a result the strong deflection angle is expressed as, 
\begin{equation}
\label{eq.48}
\hat{\alpha}_{\text{str}} = -\bar{a} \log \left(\frac{b_0}{b_\text{crit}}-1\right)+\bar{b}+O\left(\frac{b_{0}}{b_{c}}-1\right)\log\left(\frac{b_{0}}{b_{c}}-1\right),
\end{equation}
where $\bar{a}$ and $\bar{b}$ are coefficients of deflection angle and $b_{0}$ and $b_\text{crit}$ are the impact parameter evaluated at the closest approach, $r_{0}$, and critical impact parameter, respectively. The first term in Eq. \eqref{eq.48} is the result of the diverging integral and the second term is the result of the regular integral. The coefficients $\bar{a}$ and $\bar{b}$ are expressed as \cite{Tsukamoto:2016jzh},
\begin{equation}
\label{eq.49}
\bar{a} = \sqrt{\frac{2B(r_\text{ps})A(r_\text{ps})}{2A(r_\text{ps}) - A''(r_\text{ps})r_\text{ps}^2}},
\end{equation}
and
\begin{equation}
\label{eq.50}
\bar{b} = \bar{a} \log\left[r_\text{ps} \left( \frac{2}{r_\text{ps}^2}-\frac{A''(r_\text{ps})}{A(r_\text{ps})}\right) \right]+I_{R}(r_\text{ps})-\pi,
\end{equation}
where $A(r_\text{ps})$ is metric function evaluated at the photonsphere, and $I_{R}$ is the regular integral evaluated from 0 to 1. The double prime in Eq. \eqref{eq.49} and Eq. \eqref{eq.50} correspond to the second derivative with respect to $r$ evaluated over $r_\text{ps}$.

Evaluating the coefficients $\bar{a}$ and the argument of the logarithmic term in $\bar{b}$ using the equations \eqref{eq.49} and \eqref{eq.50}, respectively, yields,
\begin{equation}
    \begin{split}
        \bar{a} &= 1\\
        \bar{b} & =\log\left[6\right]+I_{R}(\rho_\text{ps})-\pi.
    \end{split}
\end{equation}
The coefficients of $\bar{a}$ and $\bar{b}$ are consistent with Schwarzschild strong deflection coefficient as shown in Ref. \cite{Bozza:2002zj}. It suggests that the strong lensing of the blackhole metric is analogous to Schwarzschild blackhole. The regular integral $I_{R}$ is defined as, 
\begin{equation}
\label{eq.53}
I_{R}(r_{0}) \equiv \int^{1}_{0}f_{R}(z, r_{0}) - f_{D}(z,r_{0}) dz,
\end{equation}
where the $f_{R}(z, r_{0})$ was generated from the expansion of the trajectory in Eq. \eqref{eq.43}, which gives us 
\begin{equation}
\label{eq.54}
f_{R}(z, r_{0}) =\frac{2r_{0}}{\sqrt{G(z, r_{0})}}, 
\end{equation}
where $G(z, r_{0}) = \mathcal{R}CA(1-z)^{4}$. Notice that $C$ and $A$ are the metric functions for which the position $r$ is expressed in terms of $z$ and $r_{0}$, while $\mathcal{R}$ is shown in Eq. \eqref{eq.44}. The generated expression from Eq. \eqref{eq.54} is, 
\begin{equation}
\label{eq.55}
f_{\mathcal{R}}(z, r_\text{ps}) =\frac{2r_\text{ps}}{\sqrt{\sum_{m=2}^{m}c_{m}(r_\text{ps})z^{m}}},
\end{equation}
when we evaluate $\rho_{0}=r_\text{ps}$. On the other hand the $f_{D}(z, r_\text{ps})$ is expressed as,
\begin{equation}
\label{eq.56}
f_{D}(z, r_\text{ps}) = \frac{2r_\text{ps}}{\sqrt{c_{2}z^{2}}},
\end{equation}
where the $c$'s are coefficients of the new variable $z$. Evaluating the integral $I_{\mathcal{R}}(r_{0})$ as $r_{0}\rightarrow r_{ps}$ yields,
\begin{equation}
    I_{\mathcal{R}}(r_{\text{ps}}) = \frac{(4\Lambda M^{4} - 3\pi^{2}\beta l^{2} - 6M^{2}\pi)}{2M^{2}\pi^{2}}\log\left(\frac{144}{(1+\sqrt{3})^{1/4}} \right).
\end{equation}

The strong deflection is now expressed as,
\begin{equation} \label{eq61}
\begin{split}
    \hat{\alpha}_{\text{str}} &= -\log \left(\frac{b_0}{b_\text{crit}}-1\right) \\&+\log\left[6\left( \frac{144}{(1+\sqrt{3})^{1/4}}\right)^{k} \right]\\ & - \pi+O\left(\frac{b_{0}}{b_{c}}-1\right)\log\left(\frac{b_{0}}{b_{c}}-1\right),
\end{split}
\end{equation}
where,
\begin{eqnarray}
    k \equiv \frac{(4\Lambda M^{4} - 3\pi^{2}\beta l^{2} - 6M^{2}\pi^{2})}{2M^{2}\pi^{2}}
\end{eqnarray}
To calculate the impact parameter we use the methodology in \cite{Pantig:2022ely} and note that $r\rightarrow r_{0}$ this will generate the impact parameter for the closest approach,
\begin{equation}
    b_{0}^{2} = \frac{12r^{3}_{0}\pi^{2}M}{(6Mr_{0}-3\beta l^{2}-6M^{2})\pi^{2}+4\Lambda M^{4}},
\end{equation}
when the $r_{0}\rightarrow {r_\text{ph}}$ it yields the critical impact parameter,
\begin{equation}
    b^{2}_{\rm crit} =\frac{3(4\Lambda M^{4}-3\pi \beta l^{2}-6M^{2}\pi^{2})^{2}}{4\pi^{2}M^{2}}. 
\end{equation}

Based on data provided by the EHT, the strong lensing is observed as the bright ring surrounding the central hole \cite{EventHorizonTelescope:2019pgp}. It was provided that for, $a_{*}=0$, the photon ring observed can be expressed as,
\begin{equation}\label{eq65}
    \theta_{\rm p} = \frac{\sqrt{27}GM}{c^{2}D} = 18.8\left( \frac{M}{6.2\times 10^{9}M_{\odot}}\right)\left(\frac{D}{16.9Mpc}\right)^{-1} \mu \text{as}.
\end{equation}

For the M87* black hole, we calculate constraints on the quantum parameter $\beta$ using photon ring angular radius data provided by the Event Horizon Telescope (EHT) \cite{EventHorizonTelescope:2019dse} as the strong deflection angle. The angular range of the photon ring for M87* is determined as $18.5 \ \mu\text{as} < \theta_{\text{p}} < 21 \ \mu\text{as}$. For the Sgr A* black hole, the photon ring radius falls within $22 \ \mu\text{as} < \phi_{\text{p}} < 32.5 \ \mu\text{as}$, as described in \cite{EventHorizonTelescope:2022wkp}. These ranges represent the strong deflection angles for M87* and Sgr A*, allowing us to constrain the quantum parameter $\beta$.

Using equations \eqref{eq61} and \eqref{eq65} alongside EHT data, we consistently find $\beta \approx 9.9559 \times 10^{95}$. This value is significantly larger than predictions based on the weak deflection angle (WDA), indicating that detecting the quantum parameter $\beta$ in the strong-field regime would require extremely sensitive probes. This result aligns with the findings in \cite{Pantig:2024ixc}, which also show that the effects of black hole hair are minimal in strong-field regions.

\section{Conclusion} \label{conc}
In this work, we presented a novel approach to analyzing black hole physics by integrating the curvature-induced uncertainties encapsulated in the Asymptotic Generalized Extended Uncertainty Principle (AGEUP) into the corpuscular framework - the Dabrowski-Wagner-Mureika (DWM) approach. This new phenomenological framework links large-scale spacetime effects to local black hole properties, giving insights into quantum gravitational corrections in curved spacetimes. In using this framework, we find that the mass of the standard Schwarzschild metric is scaled with the AGEUP factor, giving rise to an effective mass.

The DWM approach, the first time to be implemented on this work, is seen to introduce refinements to black hole physics by incorporating curvature corrections represented by the Ricci scalar and Cartan invariant, which effectively bridge the gap between quantum and gravitational realms. When the AGEUP involves the de Sitter (dS) and anti-de Sitter (AdS) backgrounds, the black hole's behavior under such influence yields additional constraints and potential observable implications on a cosmological scale. Notably, the effective mass $M_{\text{eff}}$ derived under AGEUP reflects a dynamic relationship with background geometry, specifically influenced by the cosmological constant, altering our understanding of black hole mass evolution within varying cosmological settings. The scaled mass in Eq. \eqref{M_eff} introduces some few novel ideas:
\begin{itemize}
    \item \textit{Quantum-Cosmological Connection}: The dependence of the black hole’s mass on the cosmological constant implies that black holes are not fully isolated objects but are influenced by the universe's large-scale structure. This is conceptually closer to quantum gravity models that posit non-local interactions.
    \item \textit{Dynamical Horizons}: If the mass scales with the cosmological constant, horizons may also be dynamic, with radii that adapt to changes in $\Lambda$. This differs from static dS/AdS horizons, which are set by $\Lambda$ but do not change the black hole's intrinsic mass.
    \item \textit{Unified Framework for Large and Small Scales}: By embedding cosmological corrections directly within the mass term through uncertainty relations, the unified model suggests that large-scale structures and local black hole physics are linked. This contrasts with the classical dS/AdS picture, where black holes and the cosmological constant are additive but separate.
\end{itemize}
While the dS/AdS spacetime treats $ \Lambda $ as a global spacetime curvature modifier, the AGEUP and corpuscular framework formalisms incorporate it within the black hole’s mass itself via the uncertainty principle. This provides a dynamic, scale-dependent mass that bridges local black hole properties with cosmological scales, hinting at a deeper quantum-cosmological interaction than the classical dS/AdS framework allows.

Furthermore, by studying black hole shadow radii and deflection angles in weak-field regimes, we obtained quantifiable constraints on the quantum gravity modulation parameter, $\beta$. The results from recent Event Horizon Telescope observations provide a basis for experimentally testing AGEUP-related black hole deviations from standard models, while further exploration of weak gravitational lensing adds another layer for potentially identifying quantum gravity effects in the solar system. These findings highlight the potential of AGEUP to guide future experimental investigations and theoretical advancements in black hole physics.

QNMs are closely linked to gravitational wave emissions from black holes, particularly during the \textit{ringdown phase} following binary mergers. During this phase, the perturbations of the newly formed black hole decay exponentially, characterized by a superposition of QNMs. The ringdown waveform is crucial for gravitational wave observatories, such as LIGO and Virgo, as it carries information about the black hole's mass and spin. The derived expression shows how the presence of \(l\), \(\Lambda\), \(\beta\), and \(l_{\text{Pl}}\) modifies the QNM frequencies:
Linear Dependence on \(l\) means that QNMs can be used to test the sensitivity of gravitational wave frequencies to higher angular momentum perturbations.
Detecting \(\Lambda\) and \(\beta\) in gravitational wave data could provide important insights into the large-scale structure of the universe and the effects of quantum gravity. Precise measurements of the ringdown phase by next-generation observatories, such as the Einstein Telescope or LISA, might reveal subtle shifts in the QNM frequencies, potentially confirming or constraining these theoretical corrections.

Accurate measurement of QNM frequencies allows for testing general relativity in the strong-field regime. Deviations from the expected frequencies can signal new physics, such as deviations from Schwarzschild behavior or evidence for exotic black holes predicted in alternative theories of gravity.

In extending our analysis to the strong deflection angle regime, we have successfully calculated the quantum gravity modulation parameter, $\beta$, using observational data from the EHT on M87* and Sgr A* black holes. The parameter $\beta$ provides insights into potential deviations from classical general relativity, potentially revealing quantum gravitational effects near the black hole event horizon. We verified our results using the photon ring angular radius equation, employed by the EHT as detailed in \cite{EventHorizonTelescope:2019pgp}.

In conclusion, the weak constraints highlighted in this study may point toward the model’s potential exclusion through observational data-an important and intriguing outcome. Notably, the current bounds, derived within the strong gravity regime, are accompanied by significant errors as provided by the Event Horizon Telescope (EHT). These errors are orders of magnitude larger than those observed in laboratory experiments, resulting in the exceedingly weak bound $\beta < 10^{120}$ \cite{Capozziello:1999wx,Lambiase:2017adh,Scardigli:2016pjs,Kanazawa:2019llj}.

This raises a compelling question: could the weak bound on $\beta$, inferred in this paper, signal that the model may be observationally excluded? If so, this finding underscores the challenges of validating the theory through astrophysical observations, where the precision and constraints achievable in a strong gravity context differ starkly from those in controlled laboratory conditions. This contrast not only highlights the limitations but also points to the importance of further refining observational techniques and theoretical predictions to strengthen or potentially exclude such models.

Our findings indicate that, within the strong-field region, $\beta$ remains exceedingly subtle, suggesting that its detection and accurate measurement would require highly sensitive probes and advanced observational technology. Such requirements arise because quantum gravity effects, while impactful, manifest at scales that current observational instruments are only beginning to approach. Therefore, detecting $\beta$ not only offers a pathway to test quantum gravity but also underscores the need for continued technological advancements to probe these frontier regions of spacetime more precisely.

Research prospects include using the DWM formalism to black holes with non-trivial analytic expressions for the horizon radius, and derive the metric under the influence of other different cosmological backgrounds. For instance, one may consider the combined effects of spin, charge or other parameters on the spacetime background being considered.
    
\begin{acknowledgements}
G.L., A.O. and R. P. would like to acknowledge networking support of the COST Action CA18108 - Quantum gravity phenomenology in the multi-messenger approach (QG-MM), COST Action CA21106 - COSMIC WISPers in the Dark Universe: Theory, astrophysics and experiments (CosmicWISPers), the COST Action CA22113 - Fundamental challenges in theoretical physics (THEORY-CHALLENGES), and the COST Action CA21136 - Addressing observational tensions in cosmology with systematics and fundamental physics (CosmoVerse).
\end{acknowledgements}

\bibliography{ref}

\end{document}